\documentclass[12pt]{article}
\begin{document}
\baselineskip 18pt

\newcommand{\beq}{\begin{equation}}
\newcommand{\eeq}[1]{\label{#1}\end{equation}}
\newcommand{\bea}{\begin{eqnarray}}
\newcommand{\eea}[1]{\label{#1}\end{eqnarray}}
\newcommand{\Tr}{\mbox{Tr}\,}
\newcommand{\bm}[1]{\mbox{\boldmath{$#1$}}}
\def    \lf     {\left (} \def  \rt     {\right )}
\def    \a      {\alpha} \def   \lm     {\lambda}
\def    \D      {\Delta} \def   \r      {\rho}
\def    \th     {\theta} \def   \rg     {\sqrt{g}} \def \Slash  {\, /
   \! \! \! \!}  \def      \comma  {\; , \; \;} \def       \pl
{\partial} \def         \del    {\nabla}
\def \half {1 \over 2}
\def \aad { \alpha {\dot{\alpha}}}
\def \bbd { \beta {\dot{\beta}}}
\def \bd { \dot{\beta}}
\def \ad {{\dot{\alpha}}}
\def \de {\delta_{\epsilon}}
\def \nn {\nonumber}

\begin{titlepage}
\hfill   CERN-PH-TH/2005-209, RI57/05, hep-th/0511061
\begin{center}
{\Large \bf Multitrace Deformations of Vector and Adjoint Theories and their
Holographic Duals}

\vspace{20pt}

{\large S. Elitzur$^a$, A. Giveon$^a$, M. Porrati$^{b,c}$ and E.
Rabinovici$^{a,d}$}

\vspace{12pt}

{$^a$ Racah Institute of Physics\\ the Hebrew University\\
Jerusalem, IL 91904, ISRAEL

\vspace{12pt}

$^b$ Department of Physics\\ New York University\\
4 Washington Pl.\\ New York, NY 10003, USA

\vspace{12pt}

$^c$ Scuola Normale Superiore\\ Piazza dei Cavalieri 7\\
I-56126, Pisa, ITALY
\vspace{12pt}

$^d$ PH Department, TH Unit\\ CERN\\  CH 1211 Geneva 23, SWITZERLAND}
\vspace{20pt}

\end{center}
\begin{abstract}
We present general methods to study the effect of multitrace
deformations in conformal theories admitting holographic duals in
Anti de Sitter space. In particular, we analyse the case that
these deformations introduce an instability both in the bulk AdS
space and in the boundary CFT. We also argue that multitrace
deformations of the $O(N)$ linear sigma model in three dimensions
correspond to nontrivial time-dependent backgrounds in certain
theories of infinitely many interacting massless fields on
$AdS_4$, proposed years ago by Fradkin and Vasiliev. We point out
that the phase diagram of a truly marginal large-$N$ deformation
has an infrared limit in which only an $O(N)$ singlet field
survives. We draw from this case lessons on the full
string-theoretical interpretation of  instabilities of the dual
boundary theory and exhibit a toy model that resolves the
instability of the $O(N)$ model, generated by a marginal
multitrace deformation. The resolution suggests that the
instability may not survive in an appropriate UV completion of the
CFT.
\end{abstract}
\end{titlepage}
\newpage
\section{Introduction}
Three dimensional Conformal Field Theories (CFTs) are of special
interest as they may well define macroscopic four
dimensional local theories of gravity. In addition, if the boundary theory
consists of scalar fields $\bm{\phi}$ in the fundamental
representation of a global $O(N)$ symmetry, much is known about
the system, in particular in its large $N$ limit. The emerging
structure is quite rich and includes isolated fixed points as well
as a line of fixed points and flat and unstable directions. It is
not yet known if every CFT defined on some boundary admits a
corresponding local bulk theory. In the absence of that knowledge one
can still identify several interesting features that such a
corresponding bulk theory would have, if it exists.

One such case may correspond to a limit of string theory on
$AdS_4$ in which there exists just one flat Regge trajectory with
massless particles of even spin. It was suggested in~\cite{va}
(see~\cite{va2} for a recent review) that such a theory of
massless high spin particles exists in an $AdS_4$ background. The
exact quantum description of such a system is still lacking, but
one could consider instead its CFT dual, which was conjectured
in~\cite{pk} to be a three dimensional theory of vector particles,
$\bm{\phi}$,  with a global $O(N)$ symmetry. This was based on the
identification of $O(N)$-singlet, conserved currents in the
boundary theory with the correct quantum numbers to correspond to
massless even spin particles in the bulk.  The identification (or
elimination) of the many conserved non-singlet currents was not
resolved.

In~\cite{pk} two fixed points of such three dimensional theories
of scalars were discussed: the trivial fixed point and the
non-trivial fixed point, which are known to occur in the presence
of a relevant perturbation of  the form $ g_4 (\bm{ \phi}^2)^2 $
in the large $N$ limit.  A dual mapping between these two versions
of the same bulk theory was suggested as well.  It was also
shown~\cite{gpz} that the boundary theory at the non-trivial fixed
point has all the features needed to describe a Higgs phenomenon
in the bulk. In particular, $1/N$ corrections allow all particles
with spin larger than two to obtain a mass. We review these
results in section 2. Classically, the scalar theory has also a
marginal deformation; it is generated by the operator $(\bm{
\phi}^2)^3$. For any finite {\em positive} value of the
deformation parameter $g_6$ the operator is quantum mechanically
irrelevant and this infrared free theory is only well defined for
zero renormalized value of $g_6$. Namely, for any finite value of
$g_6$ the theory can only be defined in the presence of a cutoff;
it contains only particles whose mass is of the order of the
cutoff and  thus is afflicted with all the bad properties of such
non-renormalizable theories. In the strict large $N$ limit a
window opens up,  and the theory becomes well defined for a finite
range of  $g_6$: $g_6\leq g_c$ \cite{bmb}. The behaviour of the
theory at the critical value of $g_6$ and beyond are of interest.
For example, at the critical value the scale invariance can be
spontaneously broken and a mass scale can be generated. The
effective infrared limit contains now only conserved
flavor-singlet currents.

In section 3 we review and discuss the properties of these
theories and their possible consequences for the bulk duals. This will touch
both upon the dual tensionless phase of strings and on possible quantum
resolution of classical singularities.  In section 4 we establish
a rather general effective field theory approach to usefully deal
with multi trace deformation. In section 5 we leave the safe realm
of scalar vector particles which have a well defined boundary
theory and venture into the more shady bulk description of a
system which has interesting bulk properties, including classical
space-like singularities and a strongly interacting three
dimensional CFT dual, namely the IR limit of a super Yang-Mills
theory with 16 supercharges. This theory too can be deformed by a
marginal operator of the form $(\Tr \phi^2)^3$, where $\phi$
denotes appropriate scalars of the theory. There, we also discuss
the issue and significance of possible instabilities and their
resolution in the boundary theory.

Finally, in section 6, we expand on the meaning of such instabilities in
string theory~\footnote{Other consequences of instabilities in the presence 
of a mass gap were studied in~\cite{hr}.}. 
In particular, we discuss when string theory is expected to
resolve singular bulk configurations, and when it is not.
We indicate that the instabilities in the dual theory relate essentially
to bulk configurations whose singularities are of a nature that does
not need to be resolved by string theory in a near-horizon limit.

\section{The $O(N)$ Vector Model and its Marginal Deformation}
Let us now review some known facts about the three dimensional
theory once a classically marginal operator,$(\bm{ \phi}^2)^3$, is
added~\cite{bmb}. For any finite value of $N$, the coupling $g_6$
of this operator  is infrared free quantum mechanically, as the
marginal operator gets a positive anomalous dimension already at
one loop. This implies that the theory is only well defined for
zero value of the coupling of this operator. In the
presence of a cutoff interacting particles have mass of
the order of the cutoff. At its tri-critical point the $O(N)$
model in three dimensions is described by the Lagrangian, \beq
{\cal{L}}={1\over 2}\pl_{\mu} {\bm{ \phi}} \cdot \pl^{\mu} {\bm{
\phi}} + {1\over 6N^2}g_6 (\bm{ \phi}^2)^3~, \eeq{e1} where the
fields $\bm{ \phi}$ are in the vector representation of $O(N)$. In
the limit, $N \rightarrow \infty$, \beq \beta_{g_6}=0~. \eeq{e2}
$1/N$ corrections break conformality. In the large $N$ limit,
then, $g_6$ is a modulus. There is no spontaneous breaking of the
$O(N)$ symmetry and it is instructive to write the effective
potential in terms of an $O(N)$ invariant field, \beq \sigma =
\bm{ \phi}^2~; \eeq{e3} notice that the renormalized $\sigma$ is
non-positive!~\cite{bmb}. The effective potential is: \beq
V(\sigma)= f(g_6) |\sigma|^3  \; \; , \eeq{e4} where: \beq f(g_6)=
g_c-g_6 \eeq{e5} with \beq g_c=(4\pi)^2~. \eeq{e6} The system has
various phases. For values of $g_6$ smaller than $g_c$, i.e. when
$f(g_6)$ is positive,  the system consists of $N$ massless
non-interacting $\phi$  particles. These particles do not interact
in the infinite $N$ limit; thus, correlation functions do not
depend on  $g_6$ and all the conserved currents identified
by~\cite{pk} do not aquire anomalous dimensions along that
direction. The corresponding high spin particles in the bulk are
thus still massless.

For the special value $g_6=g_c$, $f(g_6)$ vanishes and a flat
direction in $\sigma$ opens up: the expectation value of $\sigma$
becomes a modulus.  For a zero value of  this expectation value,
the theory continues to consist of $N$ massless $\phi$ fields. For
any non-zero value of the expectation value the system has $N$
 massive $\phi$ particles. All have the
same mass due to the unbroken $O(N)$ symmetry. Scale invariance is
broken spontaneously so the vacuum energy still vanishes. The
Goldstone boson associated with the spontaneous breaking of scale
invariance, the dilaton, is massless and identified as the $O(N)$
singlet field $\delta\sigma\equiv \sigma-\langle \sigma \rangle$.
All the particles are non-interacting in the infinite $N$ limit.
This theory is not conformal: in the infrared limit, it flows to
another theory containing a single, massless, $O(N)$-singlet
particle. From this singlet, one can construct even-spin conserved
currents which are now however all $O(N)$ singlets.  For larger
values of $g_6$ the exact potential is unbounded from below. The
system is unstable (in the supersymmetric case the potential is
bounded from below and the larger $g_6$ structure is similar to
the smaller $g_6$ structure~\cite{bhm}). Actually this instability
is an artifact of the dimensional regularization used above, which
does not respect the positivity of the renormalized field
$\sigma$. Still it will be useful when we discuss instabilities.
In any case a more careful analysis~\cite{bmb} shows that  the
apparent instability reflects the inability to define a
renormalizable interacting theory, all masses are of the order of
the cutoff and there is no mechanism to scale them down to low
mass values. In other words, the theory depends strongly on its UV
completion. This is summarized in Table 1.
\begin{table}[t]\centering
\caption{Marginal Perturbations of the $O(N)$ Model}
\begin{tabular}{ccccc}
$f(g_6)$   & $|\langle\sigma\rangle|$ & S.B. & masses & $V$  \\
$f(g_6) > 0$  &   0     &      No & 0  &0 \\
$f(g_6)=0$   &    0 & No &0&0 \\
$f(g_6)=0$   & $\neq 0$& Yes & Massless dilaton, $N$ & 0\\
&&& particles of equal mass & \\
$f(g_6)<0$   &  $\infty$   & Yes,  & Tachyons or masses & $-\infty$
\\ &&but ill defined & of order the cutoff  &
\end{tabular}
\end{table}
There, S.B. denotes spontaneous symmetry breaking of scale
invariance and $V$ is the vacuum energy. For $f(g_6) <0$ the
theory is unstable. Note that the vacuum energy  always vanishes
whenever the theory is well-defined.

When  $\langle\sigma\rangle \neq 0$ and the scale invariance is
spontaneously broken, one can write down the effective theory for
energy scales below $\langle\sigma\rangle$, and integrate out the
degrees of freedom above that scale. The vacuum energy remains
zero, however, and is not proportional to $\langle\sigma\rangle^3$
as might be expected naively (see~\cite{br} and references therein).

For completeness and for potential future use we note that by adding
more vector fields one has also phases in which the internal
global $O(N)$ symmetry is spontaneously broken.

An  example is  the $O(N) \times O(N)$ model~\cite{brs} with two fields in
the vector representation of $O(N)$, with Lagrangian:
\bea
{\cal{L}}&=&\pl_{\mu} {\bm{\phi}_1} \cdot \pl^{\mu}
{\bm{\phi}_1} +\pl_{\mu} {\bm{\phi}_2} \cdot \pl^{\mu}
{\bm{\phi}_2} +
\lambda_{6,0} (\bm{\phi}_1^2)^3+ \\
&& \lambda_{4,2} (\bm{\phi}_1^2)^2 (\bm{\phi}_2^2) +
\lambda_{2,4} (\bm{\phi}_1^2) (\bm{\phi}_2^2)^2 +
\lambda_{0,6} (\bm{\phi}_2^2)^3
\eea{e7}
Again the $\beta$
functions vanish in the strict $N \rightarrow \infty $ limit.
There are now two possible scales, one associated with the breakdown of a
global symmetry and another with the breakdown of scale
invariance. The possibilities are summarized by the table below:
\bea
\matrix{ & O(N)&O(N)& {\rm{scale}}&
{\rm{massless}}&{\rm{massive}} &V \cr
         & +&+&+&{\rm{all}}&{\rm{none}}&0 \cr
         &-&+&-&(N-1)\pi's, D&N,\sigma&0 \cr
&+&-&-&(N-1)\pi's, D & N,\sigma&0 \cr &-&-&-& 2(N-1)\pi's,D &
\sigma& 0 \cr }
\eea{e8}
Again, in all cases the vacuum energy
vanishes. Assume a hierarchy of scales where the scale invariance
is broken at a scale much above the scale at which the $O(N)$
symmetries are broken. One would have argued that one would have
had a low energy effective Lagrangian for the massless  pions and
dilaton with a vacuum energy given by the scale at which the
global symmetry is broken. This is not true, the vacuum energy
remains zero. This system has a critical surface, on one patch the
deep infrared theory contains only one massless particle: an
$O(N)\times O(N)$ singlet. For the other patches the deep infrared theory
is described by ${\cal O}(N)$ massless particles, most of which are not
singlets.

\section{The $AdS_4$ Dual of the Deformed $O(N)$ Model}

The detailed information on the boundary theory is not matched by as much
information about the bulk theory. As we mentioned earlier,~\cite{pk} have
conjectured that the subsector of the theory defined by
correlators of $O(N)$-singlet conserved currents is dual to one of
the theories studied by Vasiliev. These theories can be defined
only in $AdS$ spaces and contain an infinite number of massless
particles of any spin. In a sense, these theories contain one flat
``Regge trajectory.'' No statement was made about nonsinglet
currents~\footnote{Remarkably though, even at ${\cal O}(1/N)$ one can
consistently restrict the theory to the singlet sector, and
interpret the anomalous dimensions of the currents in terms of a
bulk symmetry breaking mechanism in the Vasiliev theory~\cite{gpz}.}.

In the absence of a clear identification we propose here a
modified correspondence. There are indication that the bulk theory
corresponding to the trivial Gaussian fixed point and its dual
fixed point indeed contains a flat Regge trajectory but it could
-- indeed it must~\cite{banks} -- contain many such trajectories
perhaps as many as $N^2$. These would include the non-singlet
massless fields as well as massive fields. We call the resulting
bulk theory a Fradkin-Vasiliev  system. The addition of a weakly
coupled marginal operator will not change the geometry. However
once the coupling $g_6$ reaches the critical value the situation
changes. In particular the geometry corresponding to the infrared
limit in the presence of a non zero expectation value for $\sigma$
would correspond to the pure Vasiliev system. As the boundary
theory contains a single field we expect the bulk theory to
contain a well defined geometry but with a curvature larger than
in the initial $AdS_4$. Specifically, we believe that the bulk
description of the CFT at $g_6=g_c$ goes as follows. The VEV
$\langle \sigma\rangle \neq 0$ introduces a mass scale. At
energies $E\gg |\langle \sigma\rangle|$, the boundary theory has
${\cal O}(N^2)$ conserved currents. At $E\ll |\langle
\sigma\rangle|$, the effective low-energy theory contains only the
field $\sigma$, so only currents built out of it survive.

Moreover, the boundary theory contains a-priori two dimensionless
parameters, $N$ and $g_6$. In addition it may contain a length
scale $R$ associated with the spatial part of  the world volume.
In the case of a four dimensional boundary, the corresponding
parameters are $g_{YM}^2$, $N$ and $R$. The difference is that
unlike the three dimensional system, in the large $N$ limit, the
bosonic system does not depend on the value of $g_6$ as long as it
is smaller than $g_c$; the behavior is different for $g_6=g_c$, as
described above. In the supersymmetric system at the
critical point $g_6=g_c$, scale invariance can also break spontaneously.
In this case, the far-infrared effective theory consists of two
free massless particles, the dilaton and the dilatino. The
boundary theory must thus contain appropriate additional
conserved currents. The corresponding bulk theory will consist of
additional higher spin massless states organized in
supermultiplets. Such systems make sense also for
larger values of the coupling constant, $g_6>g_c$. Once again
there is no $g_6$ dependence there. This may indicate that unlike the
four dimensional case where the bulk theory depends on three
parameters, for example, $g_s$, $N$, and $R$, in the three
dimensional boundary case the bulk theory depends on one parameter
less. A possible consequence of this is that the bulk theory has
no small curvature limit at all. In the bulk, this behavior could
be reproduced if the 4d space approaches $AdS_4$ with a radius of
curvature of order one near the boundary~\footnote{The theory
possesses a fundamental length defined as the scale at which it
becomes essentially nonlocal.}, but deep into the bulk it deviates
from it and then eventually asymptotes to another $AdS_4$ space,
with an even smaller radius of curvature. Perhaps even as small as
${\cal O}(1/N)$. Somewhat similar phenomena exist for $N=4$
SYM~\cite{gppz}. The width of the bulk transition region should be
proportional to $1/\langle\sigma\rangle$.

What about the large $g_6$ behavior? The bosonic  boundary theory
is ill defined. In particular, a configuration on the boundary
would decay in a finite time. In the bulk this means that well
defined initial data evolve into a naked singularity which reaches
the boundary in a finite time. This is an example of a bulk
singularity that is not resolved by the boundary theory. In fact
we think this could be a little misleading. The true meaning of
the instability is clear in the presence of the cutoff: the
boundary operator at finite $N$ is UV relevant, so it is  badly
behaved in the UV. In the bulk dual, a UV relevant operator
corresponds to a perturbation of $AdS_4$ that diverges near the
boundary~\footnote{Technically, this is true only for
``single-trace'' deformations, that is, for deformations
corresponding to one particle states in the bulk. On the other
hand, at finite $N$, multitrace operators mix with single trace
ones. In the bulk the dual phenomenon is that in an interacting
theory there is no sharp distinction between multi-particle and
one-particle states.}. This cannot be a localized, finite-energy
initial state, since space is changed infinitely far away from
each interior point. So, the fact that this infinite-energy
configuration evolves into a singularity is not really surprising,
and it may not contradict the hope that every finite-energy,
localized initial configuration evolves either smoothly or towards
a singularity that can be resolved by the boundary CFT.

We shall argue that the same phenomenon occurs in the case of a
more complicated boundary CFT, namely that describing the
$AdS_4\times S_7$ compactification of d=11 supergravity/M theory.
That example was considered  by Hertog and
Horowitz~\cite{hh1,hh2,hh3}. The boundary theory in that case is
more complicated, since it is the infraded limit of a 3d SYM
theory with 16 supercharges. The theory still bears some
similarity with the $O(N)$ model, since it also possesses
composite scalar operators of dimension one. We next review and
elaborate on the tools needed to monitor the modifications induced
in the bulk theory by various changes of the boundary theory.

\section{Multitrace Deformations and their Use}
In this section we rederive some known results about multitrace
deformations using a unifrom language, that is well suited to
study both vector and adjoint theories. Let us consider a CFT$_d$
with $AdS_{d+1}$ dual. An operator ${\cal O}$ of dimension
$\Delta$ (scalar, for simplicity) is associated to a bulk field
$\Sigma$. The AdS line element is
\beq
ds^2={L^2\over z^2}(dz^2 +
dx^m dx_m),
\eeq{m1}
and the boundary condition for the scalar is
\beq
\Sigma \sim \alpha z^{d-\Delta} + \beta z^\Delta.
\eeq{m2}
$\alpha$ is identified with the source for
${\cal O}$, and $\beta$ is (proportional to) the VEV of ${\cal
O}$:
\beq \alpha=J, \qquad \beta= d\langle  {\cal O}\rangle_J.
\eeq{m3}

The generating functional of the connected Green functions for
${\cal O}$, $W[J]$, is given by an action on $AdS_{d+1}$, which
depends on the field $\Sigma$ as well as other fields, such as the
metric fluctuation etc. In the case of $O(N)$ models the explicit
form of this action is still a mystery, but the boundary theory is
well known. In the case of the theory living on M-theory branes,
AdS/CFT leads us to the action of $N=1$ supergravity in 11d
compactified on $AdS_4 \times S_7$. The action must be computed
on-shell with boundary condition $\alpha=J$ for $\Sigma$.
Explictly
\beq
S[\Sigma,...]|_{\rm on-shell}\equiv S[\alpha],
\qquad  S[\alpha]|_{\alpha=J} =W[J].
\eeq{m4}
Notice that $S$
depends on $\alpha$ only, because $\beta$ becomes a fixed function
of $\alpha$ by requiring regularity inside $AdS_{d+1}$. Thanks to
the standard identity
\beq
\langle  {\cal O}\rangle_J = {\delta
W[J] \over \delta J},
\eeq{m5}
we can also write
\beq
\beta =
d{\delta S[\alpha] \over \delta \alpha}.
\eeq{m6}
Define now
$\sigma=\langle {\cal O} \rangle_J=\beta/d$. In this formula, the
VEV of ${\cal O}$ is $\beta/d$ because of the relation between
$\beta$ and the VEV of ${\cal O}$ given in Eq.~(\ref{m3})

Call $F[\sigma]$ the Legendre transform of $W[J]$. Then we have
the standard identities
\beq
F[\sigma]= W[J]-J\sigma, \qquad
\sigma={\delta W[J] \over \delta J}, \qquad J=-{\delta F[\sigma]
\over \delta \sigma}.
\eeq{m6a}
By construction, $F[\sigma]$ is
the effective action of the undeformed CFT; i.e. the generator of
1PI correlators for the operator ${\cal O}$. One can add boundary
terms to the bulk action $S$; by choosing them judiciously, and by
expressing $S$ in function of the coefficient $\beta$, one has
\beq
S[\beta]|_{\beta=\sigma}=F[\sigma].
\eeq{m6b}
Now, let us
deform the CFT by adding a multi-trace deformation $V({\cal
O})+J{\cal O}$. The effect of multitrace deformations within the
AdS/CFT duality was derived in~\cite{w,bss}. It amounts to setting
\beq
\alpha = J + V'(\beta/d).
\eeq{m7}

Equations~(\ref{m6},\ref{m7}) can be written as
\beq
\alpha= J +
V'(\sigma), \qquad \alpha=- {\delta F[\sigma] \over \delta
\sigma}.
\eeq{m10}
So, $J= -V'(\sigma) - {\delta F[\sigma]/ \delta
\sigma}$. This equation tells us that the generating functional of
the deformed theory is simply: \beq F_V[\sigma]=F[\sigma] + \int
d^dx V[\sigma]. \eeq{m11} This result is independent of the
AdS/CFT correspondence and holds for any theory admitting a large
$N$ expansion. Let us make explicit the $N$ dependence of our
formulas (in the adjoint case, to be concrete). Define ${\cal O}$
such that its VEV is ${\cal O}(N^0)$, i.e. finite in the large-$N$
limit. $W[J]$ is ${\cal O}(N^2)$ so we define $W[J]=N^2 w[J]$. The
functional integral over CFT fields $\phi$ gives 
\beq 
e^{-N^2 w[J]} = \int [d\phi] e^{-S - \int d^dx N^2 J{\cal O}}. 
\eeq{m11a}
Add a multi-trace deformation $V({\cal O})=N^2 v({\cal O})$ using
the method of~\cite{ar}, and define $F[\sigma]= N^2f[\sigma]$.
Then:
\bea
e^{-N^2 f_V[\sigma]} &=& \int [d\phi][dJ] e^{-S - \int
d^dx N^2 [J{\cal O} -J\sigma + v(\sigma)]} \nonumber \\ =\int [dJ]
e^{-N^2\{w[J] + \int d^dx [-J\sigma + v(\sigma)]\}} &=& e^{-N^2
\{f[\sigma] +\int d^dx v(\sigma)\}}.
\eea{m11b}
The last equality
holds because in the large $N$ limit the functional integral
reduces to its saddle-point approximation.

Equation~(\ref{m11}) -- or, equivalently Eq.~(\ref{m11b}) --
contains in a compact form all results on multitrace deformations
known in the literature. A similar formula was proposed in
ref.~\cite{m}. Now, let us recover a few known results, and
conclude with a few comments.
\begin{enumerate}
\item In the 3d $O(N)$ model, $F[\sigma]=-g_c \sigma^3$, $\sigma<0$~\cite{bmb}.
Then, the deformation $V(\sigma)= g_6\sigma^3$ is marginal for all
$\lambda$, and the potential is bounded below for $g_6<g_c$, as
previously announced. At $g_6=g_c$, $\sigma$ becomes a modulus.
\item Let us add a quadratic perturbation, $V=\lambda\sigma^2$,
and assume $\Delta<d/2$, as in~\cite{w,kg}. According to our
Eq.~(\ref{m11}), the effective potential $F$ is
\beq
F_V[\sigma]=
\int d^dk \tilde{\sigma}(-k)[ k^{d-2\Delta}+ \lambda ]
\tilde{\sigma}(k) + O(\sigma^3).
\eeq{m12}
Here the tilde denotes
the Fourier transform. In the UV, $k^{2\Delta-d}\gg \lambda$ and
the connected two-point function of ${\cal O}$ is
\beq
\langle
\tilde{\cal O}(k) {\cal O}(0)\rangle \sim k^{2\Delta-d}.
\eeq{m13}
This equation says that the UV dimension of ${\cal O}$ is
$\Delta$. In the IR,  $k^{d-2\Delta}\ll \lambda$, so that
\beq
\langle \tilde{\cal O}(k) {\cal O}(0)\rangle \sim {1\over
k^{d-2\Delta}+ \lambda} = {1\over \lambda} -{k^{d-2\Delta}\over
\lambda^2} + O(\lambda^{-3}).
\eeq{m14}
The first term in the
expansion is just a contact term; the second says that the IR
dimension of ${\cal O}$ is $d-\Delta$ (cfr~\cite{w,kg}). Notice
that the IR theory is nontrivial because $k^{d-2\Delta}$ is
(generically) nonlocal. Of course, a standard kinetic term for
$\sigma$, proportional to $k^2$ would give a trivial IR fixed
point. When $\Delta$ is an integer, one has to remember that the
$\sigma$ kinetic term is still nonlocal because it is of the form
$k^{2\Delta} \log k^2$.
\item In~\cite{w}, Witten gives an example of one-loop flow induced by
a double-trace perturbation. We can recover that
behavior thanks to our Eq.~(\ref{m11}).

Choose $\Delta=d/2$. Then we have, instead of Eq.~(\ref{m12}),
\beq
F_V[\sigma]= \int d^dk \tilde{\sigma}(-k)\left[ {1\over \log
(k^2/\mu^2)}+ \lambda \right] \tilde{\sigma}(k) + O(\sigma^3).
\eeq{m12a}
Next, renormalise the coupling by demanding
that the kinetic term vanishes at a constant
$k^2=\Lambda^2$, independent of $\mu$. This defines the running of
$\lambda$ by the equation
\beq
{1\over \lambda(\mu)} = -\log(\Lambda^2/\mu^2).
\eeq{m12b}
Up to obvious manipulations,
this is the same as Witten's Eq. (4.9)~\cite{w}.
\item Notice that it is always possible to create flat directions in $\sigma$
by choosing $\sigma$ constant and deforming with
$V(\sigma)=-F(\sigma)$.
\item To preserve conformal invariance, we need
\beq
V(\lambda^{\Delta}\sigma)=\lambda^d V(\sigma).
\eeq{m15}
In three dimensions this gives us back the cubic potential for
deformations of conformal weight one.
\end{enumerate}

Consider now again the perturbation $V=(g_6/3)\sigma^3$ in the
$O(N)$ model. Here $d=3$, $\Delta=1$. In our conventions, the
entire action is multiplied by $N$:
\beq
S=N\int d^3 x
[\partial_\mu \bm{ \phi} \partial^\mu \bm{ \phi} + J \bm{ \phi}^2
- J\sigma + V(\sigma)].
\eeq{m15a}
When $g_6=g_c$, a
one-dimensional moduli space appears. The modulus is the VEV
$\langle \bm{ \phi}^2\rangle=\sigma$. The equation relating
$\sigma$ to $J$ is \beq J=(g_c-g_6)\sigma^2. \eeq{m16} By keeping
the physical mass (i.e. $J$) constant, we get the RG trajectory of
ref.~\cite{bmb}, upon identifying $|\sigma|$ with the mass scale
$\mu$:
\beq
0=- {d g_6 \over d \mu}\mu + 2(g_c-g_6).
\eeq{m17}

We can also easily find the existence of a Goldstone pole
$1/p^2$ for the composite state $\bm{ \phi}^2$. Indeed, when $J$ is
nonzero $W[J]$ can be expanded in powers of derivatives as
(we specialize our results to $d=3$, $\Delta = 1$)
\beq
W[J]=N\int
d^3 x [w(J) + G(J)\partial_\mu J \partial^\mu J + ....]  .
\eeq{m18}
Here $...$ stands for higher-derivative terms. Straightforward
dimensional analysis and Eqs.~(\ref{m6a}) give
\beq w(J)=-{2\over
3} \lambda_c^{-1/2} J^{3/2}, \qquad G(J)\propto J^{-3/2}.
\eeq{m19}
The effective action $F[\sigma]$ also admits a derivative
expansion, whenever $\sigma\neq 0$:
\beq
F[\sigma]=\int d^3 x
[-(\lambda_c/3) \sigma^3  + \tilde{G}(\sigma)\partial_\mu \sigma
\partial^\mu \sigma +....].
\eeq{m20}
The kinetic term $\tilde{G}(\sigma)$ can be obtained by
explicitly performing the Legendre transformation or by
dimensional analysis. In either ways we get
$\tilde{G}(\sigma)\propto \sigma^{-1}$.

The perturbed generating functional $F_V[\sigma]$ is given by
Eq.~(\ref{m11}), so, at $g_6=g_c$, $F_V[\sigma]$ exhibits a
$1/p^2$ Goldstone pole outside the origin of the moduli space
\beq
 F_V[\sigma]={\rm constant}\times N \int d^3 x \sigma^{-1} \partial_\mu \sigma
\partial^\mu \sigma +.....
\eeq{m21}
Of course, this expansion makes sense only when
$|\partial \sigma | \ll \sigma^2$.

Let us examine next the conserved currents in the perturbed $O(N)$
model. In the free $O(N)$ model there exist an infinite number of
them:
\beq
J_{\mu_1...\mu_n}(x)= \bm{ \phi}
\stackrel{\leftrightarrow}{\partial}_{\mu_1}....
\stackrel{\leftrightarrow}{\partial}_{\mu_n} \bm{ \phi}.
\eeq{m22}
When $g_6=g_c$ and $\sigma=\langle \phi^2 \rangle \neq 0$, $\sigma$
becomes the interpolating field of a zero-mass Goldstone particle:
\beq
\langle p | \bm{ \phi}^2(x) |0\rangle = F_\sigma \exp(ip_\mu
x^\mu), \qquad p^2=0 .
\eeq{m23}
The constant $F_\sigma$ is the analog of the pion coupling strength.
Now, at leading order in
$1/N$, $\sigma$, and therefore $J$, are nonzero constants. So, the
correlators
\beq
\langle 0| J_{\mu_1...\mu_n}(x) \bm{ \phi}^2 (y) | 0
\rangle , \qquad \langle 0| J_{\mu_1...\mu_n}(x) \bm{ \phi}^2 (y)
\bm{ \phi}^2(z)| 0 \rangle,
\eeq{m24}
are one-loop integrals made of the
free, massive propagators of the fields $\phi^a$, $a=1,....,N$.
So, they are nonzero and localized. More precisely, they behave
as, e.g.
\beq
\langle 0| J_{\mu_1...\mu_n}(x) \bm{ \phi}^2 (y) | 0
\rangle \sim \exp(-J^{1/2}|x-y|), \qquad \mbox{for }|x-y| \gg J^{-1/2}.
\eeq{m25}
Because of Eq.~(\ref{m23}), this means also that the
matrix elements of currents between one- and two-Goldstone boson states are
nonzero:
\beq
\langle 0| J_{\mu_1...\mu_n}(x)|
p\rangle \neq 0, \qquad \langle 0| J_{\mu_1...\mu_n}(x)
|p_1,p_2\rangle \neq 0.
\eeq{m26}
Since $\sigma$ is the
interpolating field for the Goldstone state, this means that at
low energy $E\ll J^{1/2}$, the currents $J_{\mu_1...\mu_n}(x)$ are
\beq
J_{\mu_1...\mu_n}(x)=
A\partial_{\mu_1}....\partial_{\mu_n}\sigma (x) + B \sigma (x)
\stackrel{\leftrightarrow}{\partial}_{\mu_1}....
\stackrel{\leftrightarrow}{\partial}_{\mu_n} \sigma(x) .
\eeq{m27}
The constants $A$ and $B$ can be extracted from
Eqs.~(\ref{m24},\ref{m26}). Even before any computation, $1/N$
counting gives
\beq
A={\cal O}(1/N), \qquad B={\cal O}(1/N^2).
\eeq{m28}

In conclusion, when $g_6=g_c$, at any point outside the origin of
the moduli space, conformal invariance is broken by the VEV of
$\sigma$. At low energy $E\ll \langle \sigma \rangle$, the only
light propagating degree of freedom is the Goldstone boson
$\sigma$ itself. The dual interpretation of this phenomenon in the
bulk is that the 4d space flows from an AdS with curvature radius
$R$ to another AdS space with radius $R'\ll R$. Such flows perhaps may be
described by bulk duals of the CFT, in analogy to what was done
in $AdS_5$ in~\cite{gppz}.

It is instructive to apply the formalism developed in this section
to a cubic deformation of the IR limit of SYM with 16 supercharges
in 3d. In this case, we have a good control of the bulk dual,
while the boundary theory is hard to study directly.

\section{More on $AdS_4$ Supergravity Duals of 3d CFTs and their Instabilities}
We spent some time in analysing the $O(N)$ model because while it is very
simple as a boundary theory, yet it may shed light on a quite mysterious bulk
theory on $AdS_4$. If we want to gain more control on the bulk theory, instead,
we should look elsewhere. For instance, we could take as CFT the IR limit of
a 3d super Yang-Mills $SU(N)$ theory with 16 supercharges in the large $N$
limit.  Its holographic dual is the low-energy limit of a theory living on a
stack of M-theory branes, which is 11d supergravity compactified on
$AdS_4 \times S_7$~\cite{malda}.
The dual theory describing the correlators of
the superconformal multiplet containing the 3d stress-energy
tensor is a consistent dimensional reduction of such supergravity, namely,
$N=8$, $SO(8)$ gauged supergravity~\cite{dwn}.
The effect of multitrace deformations in this theory has been studied at
length by Hertog and Horowtiz~\cite{hh1,hh2,hh3}. In particular, in~\cite{hh1},
it was shown that a marginal, triple-trace deformation induces a big crunch
singularity in the bulk. Correspondingly, the boundary CFT is unstable: the
expectation value of the composite operator appearing in the deformation
diverges in finite time.
This result can be recovered straightforwardly using our formalism.
First of all, we have to specialize our formulas to the deformation studied
in~\cite{hh1}. The operator considered there has dimension $\Delta=1$, and it
is dual to a scalar in $AdS_4$.

The unperturbed effective action for that operator, that we
will call $F[\sigma]$ as in the previous section, is again largely determined
by dimensional analysis. More precisely, when $\sigma$ is nonzero (and large),
its kinetic term $K$ is
\beq
K[\sigma]={N^2\over 8} (
\sigma ^{-1}\partial_\mu \sigma \partial^\mu \sigma + \sigma).
\eeq{mm1}
Since the 3-d boundary CFT is strongly interacting, we cannot derive this
result by writing down the action for the adjoint scalars $\phi$ of the
super Yang-Mills theory, and
using the fact that $\sigma \propto \Tr \phi^2$. We must use instead two
properties: 1) the effective action is conformally invariant, and
so is its kinetic term. 2) $\sigma$ has dimension $\Delta=1$. By setting
$\sigma=\varphi^2$, Eq.~(\ref{mm1}) becomes the standard free action of a
conformally coupled scalar $\varphi$ (not to be confused with $\phi$!).
The linear term in $\sigma$ arises because the boundary theory lives on
$R\times S^2$, and the sphere has unit radius.
Notice that the formula we wrote makes sense only when
$\sigma^2 \gg |\partial \sigma |$;
in particular, it does not apply to the origin
of the moduli space ($\sigma=0$).

Let us add next the triple-trace perturbation $U=-N^2(f/3)\sigma^3$. Using our
general formulas, the effective action becomes $F+ \int d^3x U$, so the
perturbed potential is
\beq
V(\sigma)=N^2 {f_0-f\over 3} \sigma ^3.
\eeq{mm2}
Here we have allowed the possibility that the unperturbed action has a
(conformally invariant) potential $(f_0/3)\sigma^3$, in analogy with the
$O(N)$ case.

Consider now a cosmological solution as in~\cite{hh1,hh2,hh3},
that depends only on time.
The total energy is conserved so we can write a first integral of the
equations of motion following from the Lagrangian $K+V$~\footnote{Recall that
we use the mostly plus convention for the Lorentzian metric.}:
\beq
{1\over 8\sigma} \left({d\sigma\over dt}\right)^2 + {\sigma\over 8} +
{f_0-f\over 3} \sigma^3 = E.
\eeq{mm3}
One particular solution, obtained by setting $E=0$, is $\sigma=\alpha/\cos t$.
Substitution in the energy conservation equation gives
\beq
\alpha \sin^2 t +  \alpha \cos^2 t + {8\over 3} (f_0-f) \alpha^3 = 0,
\eeq{mm3a}
i.e. $\alpha=\sqrt{3/8(f-f_0)}$. So, when $f>f_0$ there exists a runaway
solution in which the VEV $\sigma$ of a CFT operator diverges in finite time.
The dual phenomenon in $AdS_4$ is that regular initial data evolve into a
``big crunch'' singularity that reaches the boundary in finite time~\cite{hh1}.
This is a clear cut case of singular behavior in the bulk that is not cured by
embedding supergravity in a CFT dual.

This divergence may be less dramatic than it seems. In this case the proposed
boundary theory is built out of adjoint fields. The boundary description of
the bulk configuration is a three dimensional CFT with a
non-supersymmetric yet marginal deformation of the form
$(\Tr{\phi}^2)^3$. However, using the $O(N)$ vector model as a guide,
we {\em conjecture} that such an operator is not marginal quantum mechanically,
but is actually irrelevant~\footnote{In the adjoint $N=16$ CFT, 
the triple-trace deformation mixes at
${\cal O}(1/N)$ with single-trace operators. 
Since the IR of the deformed theory
is non supersymmetric, the triple-trace will generically mix
with all operators with the same unbroken global symmetries.
These will include irrelevant operators that dominate the
OPE at short-distance. In the holographic dual, they also determine  the
near-boundary behavior of the metric.}. So, we expect
the classical  deformation to become infrared irrelevant
for any finite value of $N$.

This is a case in which stringy (i.e. ${\cal O}(1/N)$) perturbative
corrections come to the rescue, since an irrelevant operator signals
large distance modifications of the
AdS background.  The IR divergence makes the smooth initial
condition leading to a big crunch ill defined, at least in the sense that it
has infinite energy with respect to an asymptotically AdS background.

This is not the type of singularity that string theory on AdS is
supposed to resolve because  at finite $N$ the initial data
leading to a big crunch instability are not well defined near the
boundary of $AdS_4$. The dual statement is that in the presence of
a runaway solution where some of the VEVs diverge in finite time,
the CFT needs a UV completion. Moreover, the late-time behavior of
the CFT depends strongly on that completion and is therefore
non-universal. The simplest way to understand this is to notice
that in order to define the CFT, one needs to introduce a UV
cutoff $\Lambda$, even when the theory is finite ~\footnote{A
simple example is the $O(N)$ model when $ g_6$ is larger than the
critical value. There, a careful treatment of the theory already
at $N=\infty$ shows that in
 the presence of a cutoff all masses are of the order of the cutoff, so that
the theory does not possess a universal low energy limit.}.

In the case of strings/M-theory, this UV cutoff is quite
physical: it is the string/M-theory mass scale. The effect of the cutoff
manifests itself in $F[\sigma]$ through higher-dimensional operators,
weighted by inverse powers of $\Lambda$. Thus, the complete potential of
$V$ is not as in Eq.~(\ref{mm2}); rather, it schematically reads
\beq
V(\sigma)= N^2 {f_0-f\over 3} \sigma ^3
+ \sum_{n=4}^\infty c_n \Lambda^{3-n}\sigma^n.
\eeq{mm4}
The coefficients $c_n$ can be of order $N^2$ or less. In any case, the effect
of the higher-dimensional operators cannot be neglected when $\sigma$ is so
large that for some $n$
$N^2(f_0-f) \sigma ^3 \sim c_n \Lambda^{3-n}\sigma^n$.

The system needs to be modified in the UV and in that case
one simply does not know  what really does happen at late times when the
singularity approaches the boundary.

Finally, we should point out that in some cases
the UV modification can be quite simple
and involve no exotic
physics. It could be as simple as an extra massive degree of freedom.

To see that, let us consider again
the deformed $O(N)$ model we studied in sections 2 and 3
and add to it a single  massive scalar $S$. The Lagrangian density is
\beq
L={N\over 2}\left[\partial_m {\bm{\phi}} \partial^m {\bm{\phi}} +
\partial_m S \partial^m S +
\alpha M^{3/2}{\bm{\phi}}^2S +{1\over 2}M{\bm{\phi}}^2S^2 +
{\beta\over 2}M ({\bm{\phi}}^2)^2 +{\gamma\over 2} M^2 S^2\right].
\eeq{uv1}
Here $M$ is a mass parameter and $\alpha,\beta,\gamma$ are arbitrary, positive,
dimensionless  constants. The potential is obviously bounded below.

At energy scales $E\ll M$ we can integrate
out the field $S$ using its equations of motion. The contribution from the
kinetic term is negligible so we have
\beq
\alpha M^{3/2}{\bm{\phi}}^2 + (M{\bm{\phi}}^2 + \gamma M^2)S=0.
\eeq{uv2}
Substituting into eq.~(\ref{uv1}) we get a potential for ${\bm{\phi}}$
\beq
V= N\left[ {\beta\over 2}M({\bm{\phi}}^2)^2 -{\alpha^2 M^3 ({\bm{\phi}}^2)^2
\over 2M {\bm{\phi}}^2 + 2\gamma M^2}\right].
\eeq{uv3}
The term proportional to $({\bm{\phi}}^2)^2$ in this potential can be canceled
by setting $\alpha^2/\gamma=\beta$. Then, the potential starts with the
dimension-3 operator $({\bm{\phi}}^2)^3$.
Precisely:
\beq
V= N\left[ -{\alpha^2 M\over 2\gamma}({\bm{\phi}}^2)^2
\sum_{n=1}^\infty \left(-{{\bm{\phi}}^2
\over \gamma M}\right)^n\right].
\eeq{uv4}
Define $g=\alpha^2/2\gamma^2$. Then, in the limit
$\gamma\rightarrow \infty$, $\alpha\rightarrow \infty$,
$g=\,$constant, all irrelevant terms vanish as inverse powers of $\gamma$.

Notice that the UV complete theory is always stable, even when $g>g_c$.
The UV completion is recovered by ``integrating in'' the field $S$. Unlike the
case of $N=1$ superpotentials, here the procedure is highly non unique.

This toy example suggests us also an alternative description of the resolution
of the singularity in the adjoint model. In that case, the UV theory is most
likely strongly interacting. Thus, terms which may seem unstable when
extrapolated from weak coupling could actually represent a stable potential,
when evaluated appropriately at strong coupling around a UV fixed point.
So, the large-${\bm{\phi}}^2$ instability cannot be used as a sure
diagnostic to invalidate the theory.

\section{Discussion/Conclusions}
We conclude with some general remarks on the meaning
of instabilities occurring both in the bulk and in the
dual boundary CFT, and whether string theory should cure those instabilities.
The argument could be posed in the following manner.
Assume one is somehow
given a bulk theory containing a singularity and its corresponding CFT
 boundary dual, also singular or unstable.
What is the interpretation of the instability in a complete string
theoretical picture?

 One may think that an instability of the boundary CFT is a disaster.
In fact one can imagine instabilities in several forms. They could
 correspond to non-positive definite, relevant operators in the boundary theory
 which  destabilize a flat direction. They could be marginal unstable
 deformations, and they could also come in the form of irrelevant, unstable
 boundary operators.  The relevant and marginal operators need to be protected
 so that  they do not become irrelevant as the
 coupling increases. All these cases have in common the feature that they
 indeed destabilize the boundary theory. Moreover, such deformations
 correspond to introducing various forms of repulsive forces among the branes
 whose near-horizon limit generated the bulk
 geometry.  Such configurations will be unstable in the bulk and the branes
 will search for a new equilibrium configuration. Generically they may do so by
 fleeing to infinity, perhaps flat 10d space.  These are indeed unstable
 configurations {\em ab initio}
and it is not one of the tasks of string theory on
AdS to resolve them. Differently said, these pathologies signal instabilities
of the near-horizon limit of string theory, not of string theory
{\em per se}.~\cite{sw,mm,kpr}.

If we examine more closely the different types of instabilities we
find that they manifest themselves in the following manner. In the
relevant case, the instability may emerge at some time, say $t=0$,
deep inside the bulk and then spread out to the boundary along a
null geodesic. On the other hand an irrelevant boundary
instability generates already from the ``start" at $t=0$ a greatly
deformed bulk metric near the boundary. This is the particular
case we have discussed in this note and it corresponds to an
unstable bulk configuration which string theory need not resolve.
This observation still leaves open the question of what would be
the appropriate diagnostic for a real failure of non-perturbative
string theory to resolve a singularity. For the time being, all
instabilities exhibited in the literature signal at most a
sickness of the near-horizon limit. On the other hand, once a
stable CFT dual to an AdS background exists, then one knows that
all bulk singularities are indeed resolved. For instance, the
unstable deformation of the $O(N)$ model at $g_6>g_c$ can be
stabilised by the simple UV completion we described at the end of
last section. Once this is done, the bulk singularity no longer
reaches the boundary. The same phenomenon happens in the CFT dual
to 11d supergravity on $AdS_4\times S_7$ we studied in section 5:
when the unstable potential is stabilised by irrelevant operators,
the singularity does not hit the boundary but evolves instead into
a giant black hole~\cite{hh3}. 

Finally we may ask ourselves if there exist 
universal properties shared by all resolutions of the instabilities of the
low-energy sector of the dual boundary field theory. When the instability is
resolved by introducing irrelevant operators the theory may seem to lose 
its predictive power. On the other hand there is a possible universal
signature of stabilization: a boundary theory
consisting of one or several stationary points, stable as well as  
unstable, may enable one to resolve a cosmological singularity. 
The story may go always as follows: near the
unstable extrema of the potential, the bulk theory will effectively 
describe a big bang or big crunch, but that potentially singular behavior
will eventually change into another one, which in the boundary theory will be
described as a relaxation towards one of its true stable minima. 
\subsection*{Acknowledgments}
We thank O. Aharony, I. Klebanov, H. Neuberger and A. Polyakov for
useful discussions. M.P. and E.R.
would like to thank the KITP at UCSB for hospitality. S.E., A.G. and M.P.
would like to thank CERN for hospitality. This work is
supported in part by the American-Israeli
Bi-National Science Foundation, the Israel Research Foundation, the European
Union grant MRTN-CT-2004-512194, and DOE grant
DE-FG02-90ER40560. M.P. is supported in part by NSF grant PHY-0245068, and by
a Marie Curie chair, contract MEXC-CT-2003-509748 (SAG@SNS).


\begin{thebibliography}{00}
\bibitem{va}
E.~S.~Fradkin and M.~A.~Vasiliev,
%``Cubic Interaction In Extended Theories Of Massless Higher Spin Fields,''
Nucl.\ Phys.\ B {\bf 291}, 141 (1987);
  %``On The Gravitational Interaction Of Massless Higher Spin Fields,''
  Phys.\ Lett.\ B {\bf 189}, 89 (1987);
\bibitem{va2}
M.~A.~Vasiliev,
%``Higher spin gauge theories in any dimension,''
Comptes Rendus Physique {\bf 5}, 1101 (2004)
[arXiv:hep-th/0409260].
\bibitem{pk}
I.~R.~Klebanov and A.~M.~Polyakov,
  %``AdS dual of the critical O(N) vector model,''
  Phys.\ Lett.\ B {\bf 550}, 213 (2002)
  [arXiv:hep-th/0210114].
\bibitem{gpz}
  L.~Girardello, M.~Porrati and A.~Zaffaroni,
  %``3-D interacting CFTs and generalized Higgs phenomenon in higher spin
  %theories on AdS,''
  Phys.\ Lett.\ B {\bf 561}, 289 (2003)
  [arXiv:hep-th/0212181].
\bibitem{bmb}W.~A.~Bardeen, M.~Moshe and M.~Bander,
%``Spontaneous Breaking Of Scale Invariance And The Ultraviolet Fixed Point In
%O(N) Symmetric (Phi**6 In Three-Dimensions) Theory,''
Phys.\ Rev.\ Lett.\  {\bf 52}, 1188 (1984).
\bibitem{hr}
J.~L.~F.~Barbon and C.~Hoyos,
%``AdS/CFT, multitrace deformations and new instabilities of nonlocal  string
%theories,''
JHEP {\bf 0401}, 049 (2004)
[arXiv:hep-th/0311274].
\bibitem{bhm}
W.~A.~Bardeen, K.~Higashijima and M.~Moshe,
%``Spontaneous Breaking Of Scale Invariance In A Supersymmetric Model,''
  Nucl.\ Phys.\ B {\bf 250}, 437 (1985).
\bibitem{br}
D.~S.~Berman and E.~Rabinovici,
%``Supersymmetric gauge theories,''
arXiv:hep-th/0210044.
\bibitem{brs}
E.~Rabinovici, B.~Saering and W.~A.~Bardeen,
%``Critical Surfaces And Flat Directions In A Finite Theory,''
Phys.\ Rev.\ D {\bf 36} (1987) 562.
\bibitem{banks} T. Banks, private communication.
\bibitem{hh1}
  T.~Hertog and G.~T.~Horowitz,
  %``Towards a big crunch dual,''
  JHEP {\bf 0407}, 073 (2004)
  [arXiv:hep-th/0406134].
\bibitem{hh2}
T.~Hertog and G.~T.~Horowitz,
  %``Designer gravity and field theory effective potentials,''
  Phys.\ Rev.\ Lett.\  {\bf 94}, 221301 (2005)
  [arXiv:hep-th/0412169].
\bibitem{hh3}
T.~Hertog and G.~T.~Horowitz,
  %``Holographic description of AdS cosmologies,''
  JHEP {\bf 0504}, 005 (2005)
  [arXiv:hep-th/0503071].
\bibitem{w}
E.~Witten,
%``Multi-trace operators, boundary conditions, and AdS/CFT correspondence,''
arXiv:hep-th/0112258.
\bibitem{bss} M.~Berkooz, A.~Sever and A.~Shomer,
%``Double-trace deformations, boundary conditions and spacetime
%singularities,''
JHEP {\bf 0205}, 034 (2002) [arXiv:hep-th/0112264];  A.~Sever and A.~Shomer,
%``A note on multi-trace deformations and AdS/CFT,''
JHEP {\bf 0207}, 027 (2002) [arXiv:hep-th/0203168].
\bibitem{ar}
D.~J.~Amit and E.~Rabinovici,
%``Breaking Of Scale Invariance In Phi**6 Theory: Tricriticality And Critical End Points,''
Nucl.\ Phys.\ B {\bf 257}, 371 (1985).
\bibitem{m} W.~Muck,
%``An improved correspondence formula for AdS/CFT with multi-trace operators,''
Phys.\ Lett.\ B {\bf 531}, 301 (2002) [arXiv:hep-th/0201100].
\bibitem{kg}
S.~S.~Gubser and I.~R.~Klebanov,
%``A universal result on central charges in the presence of double-trace
%deformations,''
arXiv:hep-th/0212138.
\bibitem{gppz}
L.~Girardello, M.~Petrini, M.~Porrati and A.~Zaffaroni,
%``Novel local CFT and exact results on perturbations of N = 4 super Yang-Mills from AdS dynamics,''
JHEP {\bf 9812}, 022 (1998) [arXiv:hep-th/9810126].
\bibitem{malda}
J.~M.~Maldacena,
  %``The large N limit of superconformal field theories and supergravity,''
  Adv.\ Theor.\ Math.\ Phys.\  {\bf 2}, 231 (1998)
  [Int.\ J.\ Theor.\ Phys.\  {\bf 38}, 1113 (1999)]
  [arXiv:hep-th/9711200].
\bibitem{dwn}
B.~de Wit and H.~Nicolai,
%``N=8 Supergravity,''
Nucl.\ Phys.\ B {\bf 208}, 323 (1982).
\bibitem{sw}
N.~Seiberg and E.~Witten,
  %``The D1/D5 system and singular CFT,''
  JHEP {\bf 9904}, 017 (1999)
  [arXiv:hep-th/9903224].
\bibitem{mm}
J.~Maldacena and L.~Maoz,
  %``Wormholes in AdS,''
  JHEP {\bf 0402}, 053 (2004)
  [arXiv:hep-th/0401024].
\bibitem{kpr}
M.~Kleban, M.~Porrati and R.~Rabadan,
  %``Stability in asymptotically AdS spaces,''
JHEP {\bf 0508}, 016 (2005)
[arXiv:hep-th/0409242].
\end{thebibliography}
\end{document}